\documentclass[conference]{IEEEtran}
\IEEEoverridecommandlockouts
\usepackage{cite}
\usepackage{amsmath,amssymb,amsfonts}
\usepackage{algorithmic}
\usepackage{graphicx}
\usepackage{textcomp}
\usepackage{xcolor}
\usepackage{multirow}

\usepackage{algorithm}
\usepackage{algorithmic}
\usepackage{subfigure}

\def\BibTeX{{\rm B\kern-.05em{\sc i\kern-.025em b}\kern-.08em
    T\kern-.1667em\lower.7ex\hbox{E}\kern-.125emX}}
    
\usepackage[bottom=1in,top=0.75in,left=0.626in,right=0.626in]{geometry}
\begin{document}

\title{Real-Time High-Resolution Pedestrian Detection in Crowded Scenes via Parallel Edge Offloading}

\author{Hao Wang, Hao Bao, Liekang Zeng, Ke Luo, Xu Chen\\
\IEEEauthorblockA{School of Computer Science and Engineering, Sun Yat-sen University, Guangzhou, China\\
Email: \{wangh688, baoh8, zenglk3, luok7\}@mail2.sysu.edu.cn, chenxu35@mail.sysu.edu.cn}
}

\maketitle

\begin{abstract}
To identify dense and small-size pedestrians in surveillance systems, high-resolution cameras are widely deployed, where high-resolution images are captured and delivered to off-the-shelf pedestrian detection models. However, given the highly computation-intensive workload brought by the high resolution, the resource-constrained cameras fail to afford accurate inference in real time. To address that, we propose \textsc{Hode}, an offloaded video analytic framework that utilizes multiple edge nodes in proximity to expedite pedestrian detection with high-resolution inputs. Specifically, \textsc{Hode} can intelligently split high-resolution images into respective regions and then offload them to distributed edge nodes to perform pedestrian detection in parallel. A spatio-temporal flow filtering method is designed to enable context-aware region partitioning, as well as a DRL-based scheduling algorithm to allow accuracy-aware load balance among heterogeneous edge nodes. Extensive evaluation results using realistic prototypes show that \textsc{Hode} can achieve up to 2.01× speedup with very mild accuracy loss.
\end{abstract}


\section{Introduction}
In recent years, many cameras have been deployed in some key places with high pedestrian traffic to monitor pedestrians for public safety reasons \cite{bib1}. To identify pedestrians appearing in videos, pedestrian detection models (e.g., Faster R-CNN \cite{ren2017faster}, YOLOv5 \cite{glenn_jocher_2021_5563715}) are widely used. However, the mainstream pedestrian detection models are typically computation-intensive, presenting significant conflict with the cameras' limited computing capability. Running pedestrian detection models on cameras is slow and difficult to meet the demand of real-time processing.

To solve this problem, some existing works \cite{zhang2017live,du2020server} have proposed uploading videos to a powerful cloud server for fast inference. However, uploading videos to a remote cloud server requires a long network transmission delay and may also risk at significant privacy leakage. To address that, some literature \cite{li2021mass,ren2018distributed,hanyao2021edge} have proposed offloading videos to some nearby edge nodes (e.g., 5G mobile edge computing (MEC) edge nodes and IoT gateway nodes) for efficient inference to solve the problem of long network transmission delay and achieve more privacy-friendly local processing in proximity.

\begin{figure}
    \centering
	\subfigure[An example of crowded scenes.]{
		\begin{minipage}[t]{0.5\linewidth}
			\centering
			\includegraphics[scale=0.215]{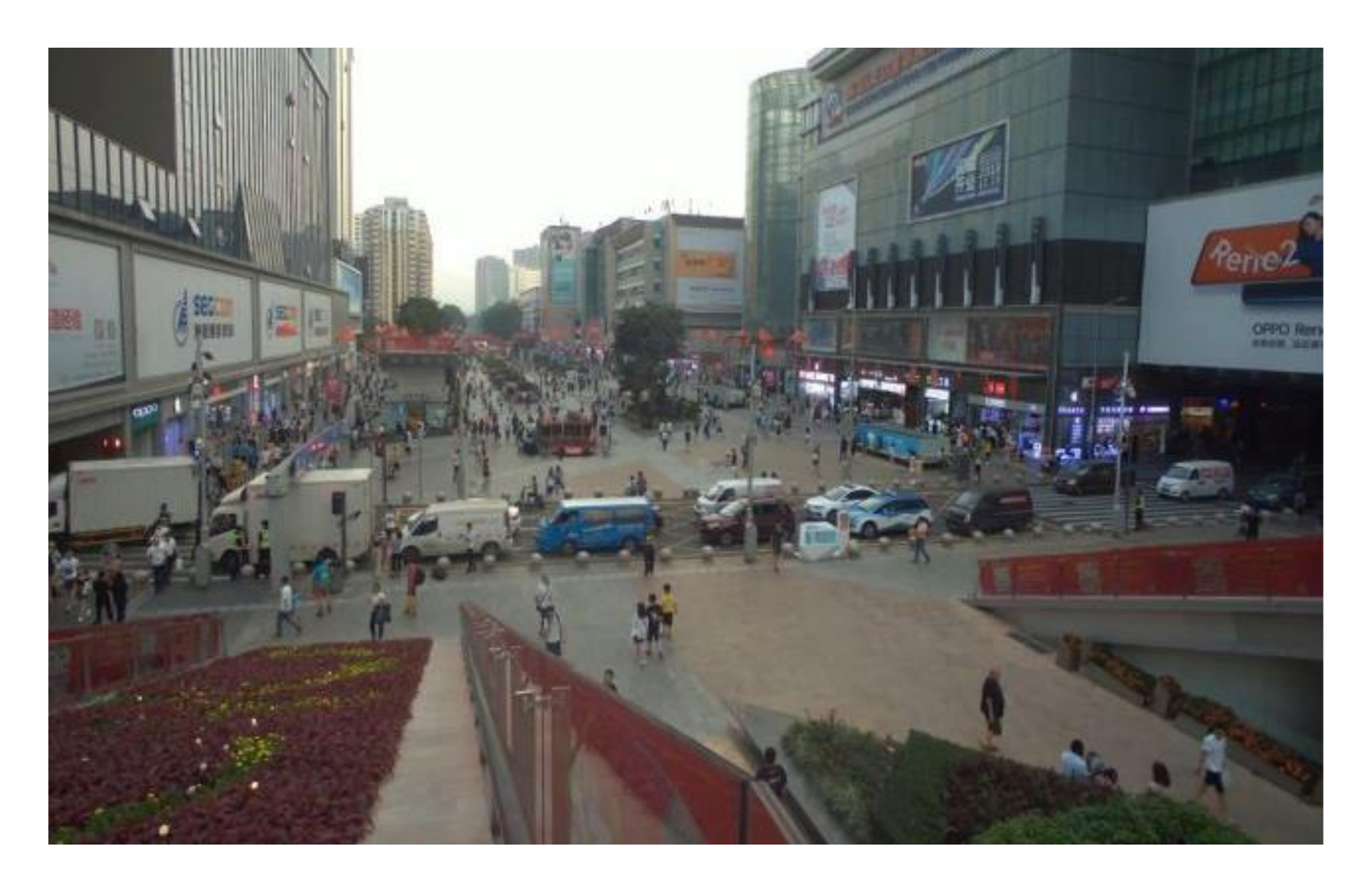}
		\end{minipage}
		\label{fig:crowded}
	}%
	\centering
	\subfigure[An illustration of filtering.]{
		\begin{minipage}[t]{0.5\linewidth}
			\centering
			\includegraphics[scale=0.215]{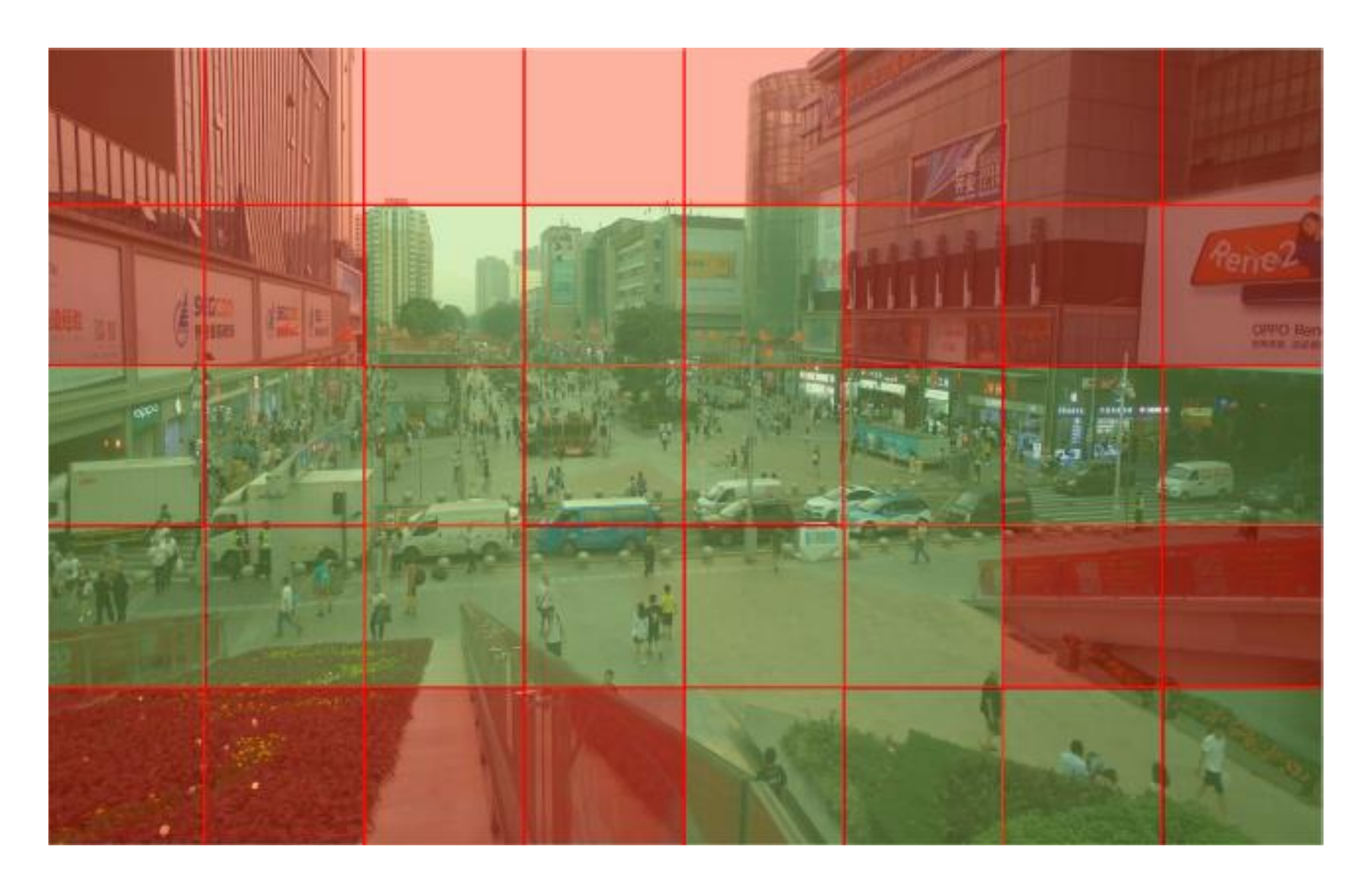}
		\end{minipage}
		\label{fig:flitering}
	}
	\caption{(a) shows an example of crowded scenes. (b) shows the filtering in \textsc{Hode}.}
	\label{fig:fig2}
\vspace{-0.3cm}
\end{figure}

However, for crowded scenes, it is non-trivial to perform fast pedestrian detection on edge nodes. In order to accurately identify pedestrians in crowded scenes, high-resolution images are employed. As shown in Fig.~\ref{fig:crowded}, this 4K image contains over 800 pedestrians. For such crowded scenes, if a low resolution is used, then many small pedestrians will have only tens or even several pixels, which seriously reduces the accuracy of pedestrian detection. Therefore, pedestrian detection in crowded scenes requires high resolution. However, performing high-resolution pedestrian detection on resource-constrained edge nodes is slow and may take several seconds.

\begin{figure}[t]
    \begin{minipage}[t]{0.2\textwidth}
        \includegraphics[width=1\textwidth]{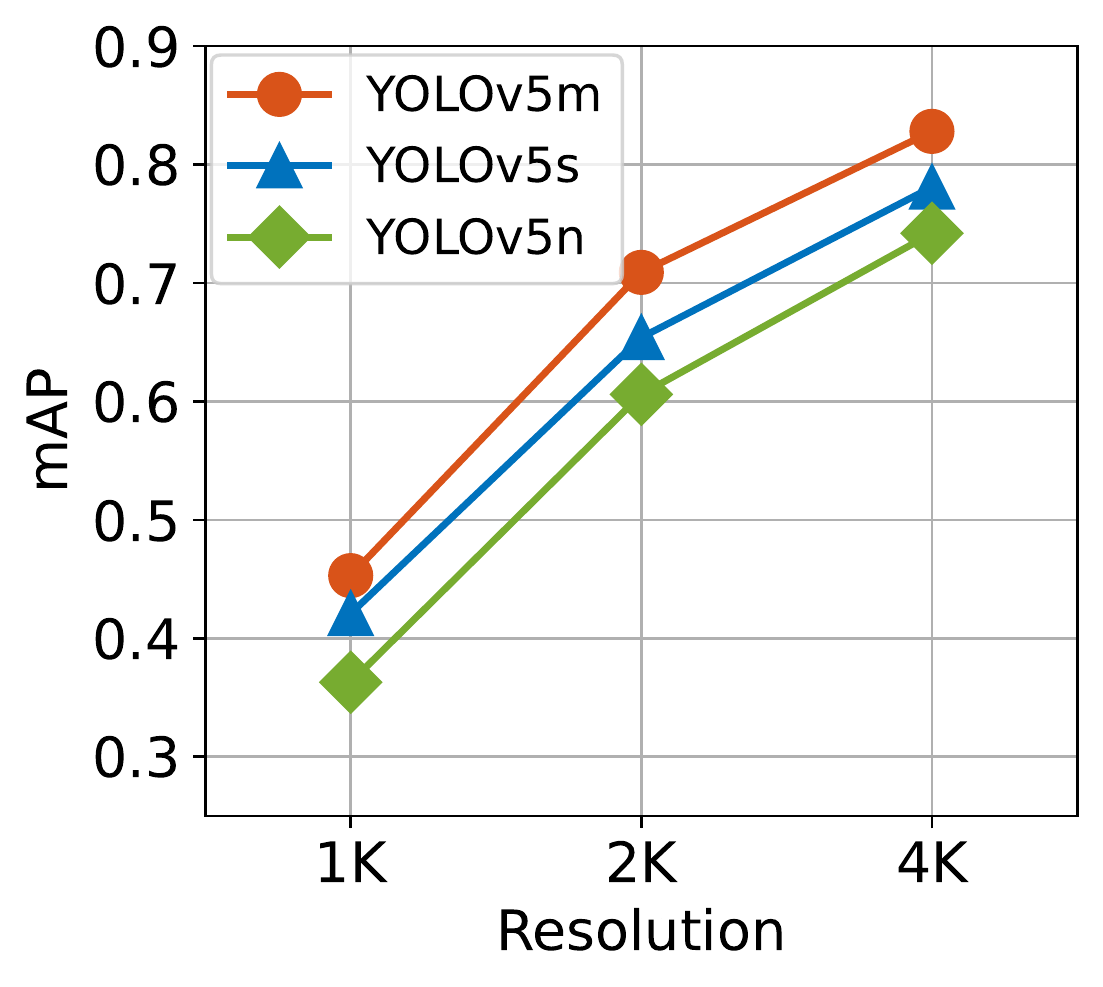}
        \caption{mAP at different resolutions on YOLOv5.}
        \label{fig:mAP_Resolution}
    \end{minipage}
    {}
    \begin{minipage}[t]{0.28\textwidth}
        \includegraphics[width=1\textwidth]{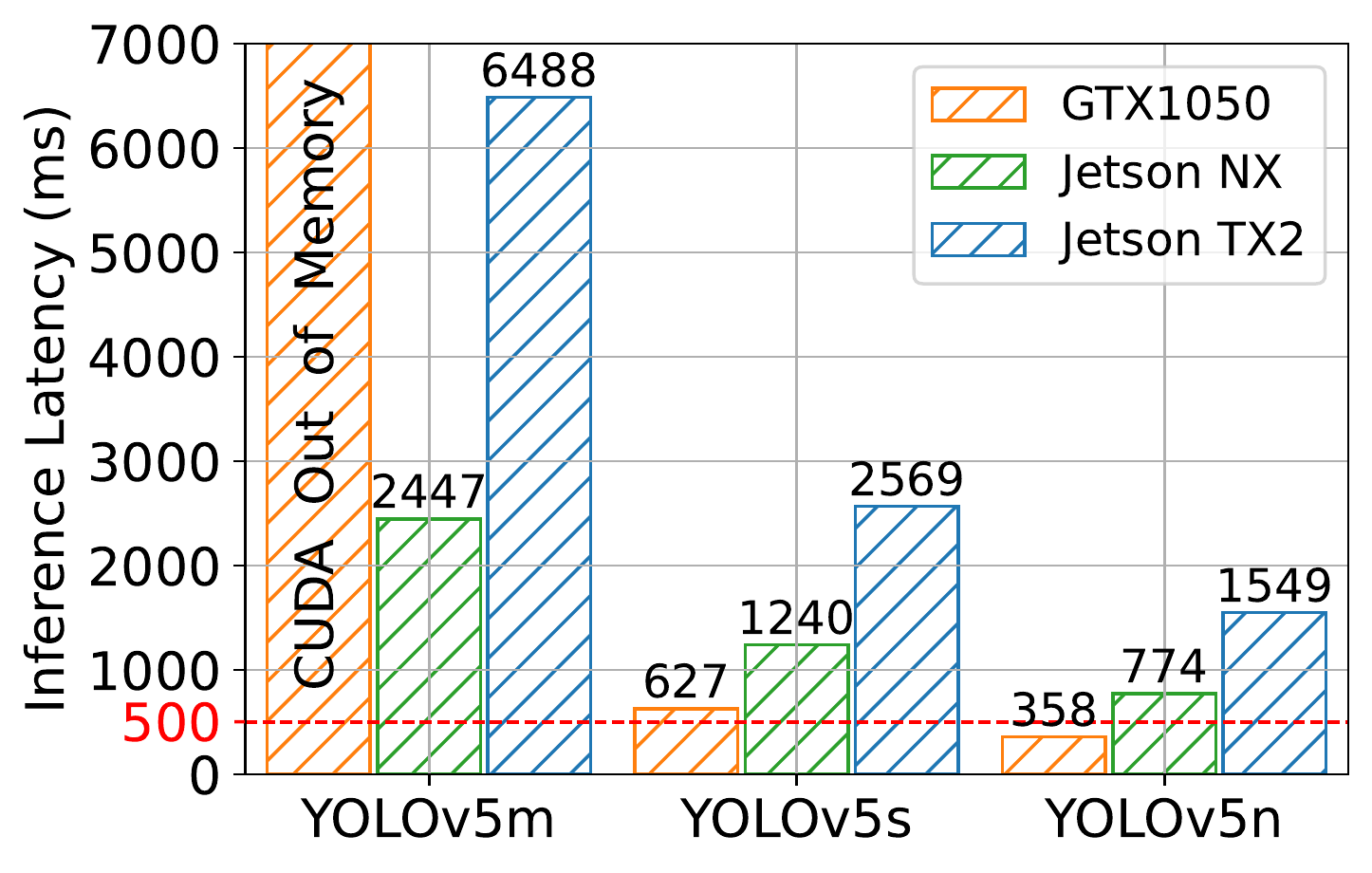}
        \caption{Inference latency of 4K images on different devices.}
        \label{fig:Latency_Device}
        \end{minipage}
\vspace{-0.3cm}
\end{figure}

To prove this, we conduct some experiments on PANDA dataset \cite{PANDA} (a high-resolution video dataset with high pedestrian traffic). We put PANDA dataset into YOLOv5 at different resolutions and then obtain the mean average precision (mAP) of pedestrian detection. As shown in Fig.~\ref{fig:mAP_Resolution}, the mAP of pedestrian detection is proportional to the input resolution. If we use 1K resolution, the final mAP is only about 0.4, because there are many small pedestrians that are not successfully identified. Therefore, for crowded scenes, high-resolution (e.g., 4K) images are required to obtain a satisfactory accuracy. 

However, if we put 4K images into pedestrian detection models, the inference latency is intolerable and may even trigger out-of-memory problems. As shown in Fig.~\ref{fig:Latency_Device}, the inference latency on some devices exceeds 500ms and even reaches several seconds, which is intolerable for the demand of real-time. So, we need some methods to speed up high-resolution pedestrian detection. However, the acceleration methods proposed by some existing researches \cite{li2021mass,ren2018distributed,hanyao2021edge} are aimed at low-resolution object detection. Other existing researches \cite{kang2017neurosurgeon,zhao2018deepthings,eshratifar2019jointdnn,zeng2020coedge,chen2020knowledge} split the layers of deep neural networks and accelerate the inference in a model-parallel manner. Nonetheless, these methods do not take into account the possibility of accelerating high-resolution pedestrian detection by skipping some regions where there are no pedestrians.

Some existing works \cite{jiang2021flexible,zhang2021elf} have proposed to split the high-resolution images into some small regions and then skip the small regions containing only the background and perform pedestrian detection on the small regions with pedestrians. However, since the pedestrians are moving, predicting whether each region is background or not is challenging. Remix \cite{jiang2021flexible} uses a simple rule to skip background regions, but cannot accurately predict the background regions. Elf \cite{zhang2021elf} uses an attention-based long short-term memory (LSTM) network to predict the location of each pedestrian before pedestrian detection and perform pedestrian detection on these locations. But Elf can be slow for crowded scenes with hundreds of pedestrians due to hundreds of attention-based LSTM inferences before pedestrian detection.

To address these limitations, in this paper, we propose HODE, an edge-assisted video analytics framework that is able to judiciously recognize background regions and leverage parallel edge offloading (e.g., utilize the MEC edge nodes and IoT gateway nodes in proximity for collaborative computing) to expedite high-resolution pedestrian detection. Specifically, to accurately and quickly predict whether each region is background or not, we propose a spatio-temporal flow filtering method to predict and filter out the small regions without pedestrians before pedestrian detection. For example, in Fig.~\ref{fig:flitering}, the green regions are predicted to have pedestrians and the red regions are predicted to be the background. We only need to perform pedestrian detection for the green regions, which can accelerate high-resolution pedestrian detection.

In addition, to further accelerate high-resolution pedestrian detection, cameras can assign small regions to some nearby edge nodes to perform pedestrian detection in parallel. However, the detection completion time of a high-resolution image depends on the slowest completion time of all small regions (i.e., the straggler). To alleviate that, we propose an accuracy-aware deep reinforcement learning (DRL) based load-balanced scheduling algorithm, which can decide a region dispatching strategy to well balance the workload among multiple heterogeneous edge nodes while achieving high accuracy of pedestrian detection.

The main contributions of this paper are as follows.
\begin{itemize}
\item We propose \textsc{Hode}, a real-time high-resolution pedestrian detection framework at the edge. \textsc{Hode} employs a lightweight spatio-temporal flow filtering method to predict and filter out small regions without pedestrians based on spatial and temporal correlations to speed up high-resolution pedestrian detection. 

\item We propose an accuracy-aware DRL-based load-balanced scheduling algorithm. This scheduling algorithm considers the computing capability heterogeneity of multiple edge nodes and aims at alleviating the straggler problem for achieving fast pedestrian detection with high accuracy.

\item We implement \textsc{Hode} on varying heterogeneous edge devices. Evaluation results show that \textsc{Hode} can achieve 2.01× speedup on high-resolution pedestrian detection with less than 1\% accuracy sacrifice compared with the ordinary parallel edge offloading scheme. In addition, our spatio-temporal flow filtering can accurately and quickly filter out regions without pedestrians before pedestrian detection.
\end{itemize}

\section{System Design}
\begin{figure}[t]
\centering
\includegraphics[scale=0.468]{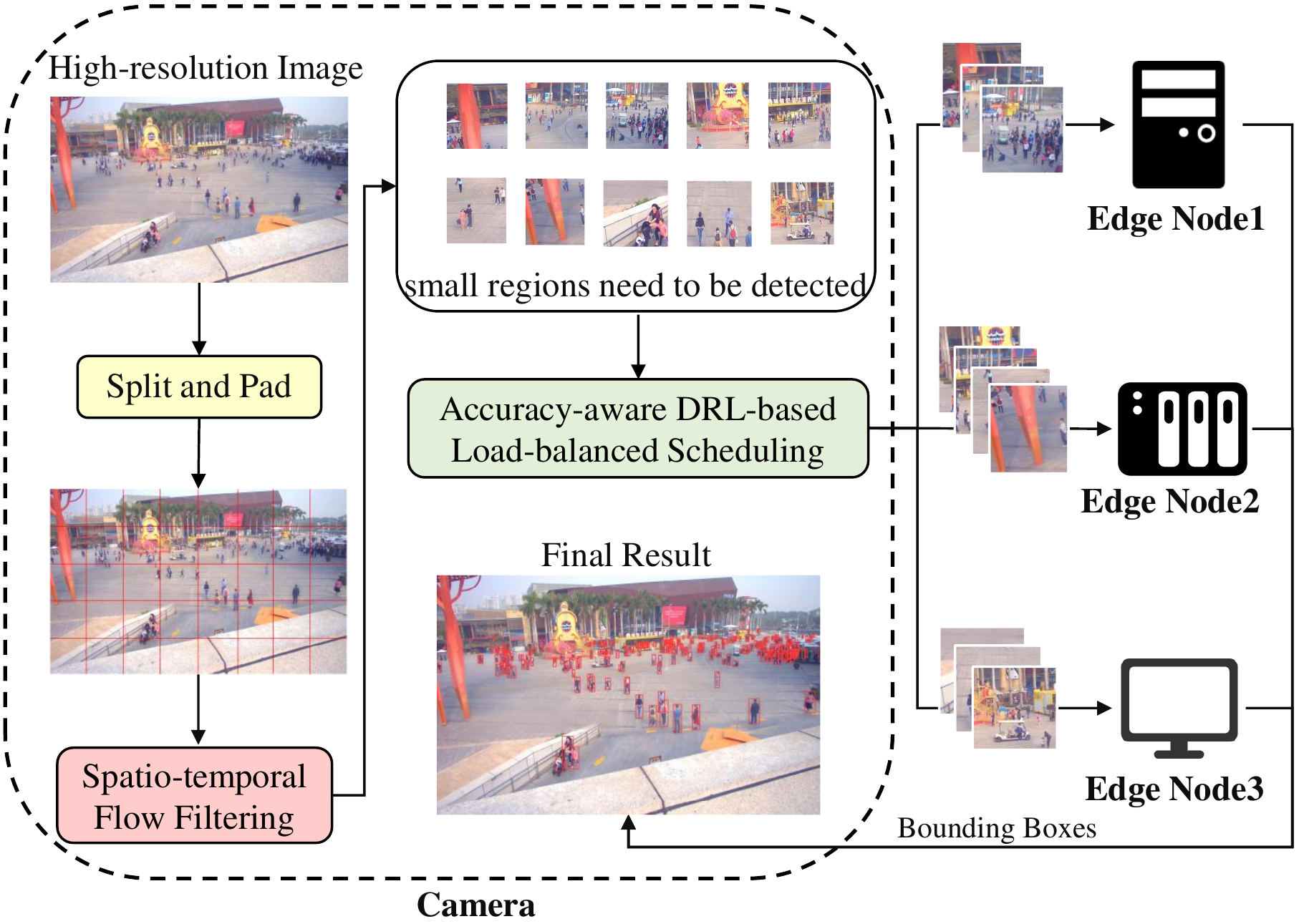}
\caption{\textsc{Hode} system overview.}
\label{fig:System}
\vspace{-0.3cm}
\end{figure}

The design of \textsc{Hode} is shown in Fig.~\ref{fig:System}. A smart camera first splits a high-resolution image into some small regions by a given size (e.g., 512 × 512). However, splitting will cause the pedestrians on the split lines to be split into two parts, which may cause the detection to fail. To solve this problem, \textsc{Hode} pads the split regions so that the split regions will cover part of each other. 
\iffalse
The benefits of splitting high-resolution images are as follows.
\begin{itemize}
\item Splitting increases the opportunities for parallel inference. As shown in Fig.~\ref{fig:System}, the camera can send small regions to multiple edge nodes. These edge nodes can perform parallel inference on small regions from the same high-resolution image.
\item Splitting helps filter out unnecessary pedestrian detection. As shown in Fig.~\ref{fig:flitering}, splitting helps filter out regions where there are no pedestrians, thus reducing unnecessary inference.
\end{itemize}
\fi

\begin{figure}[t]
\centering
\includegraphics[scale=0.25]{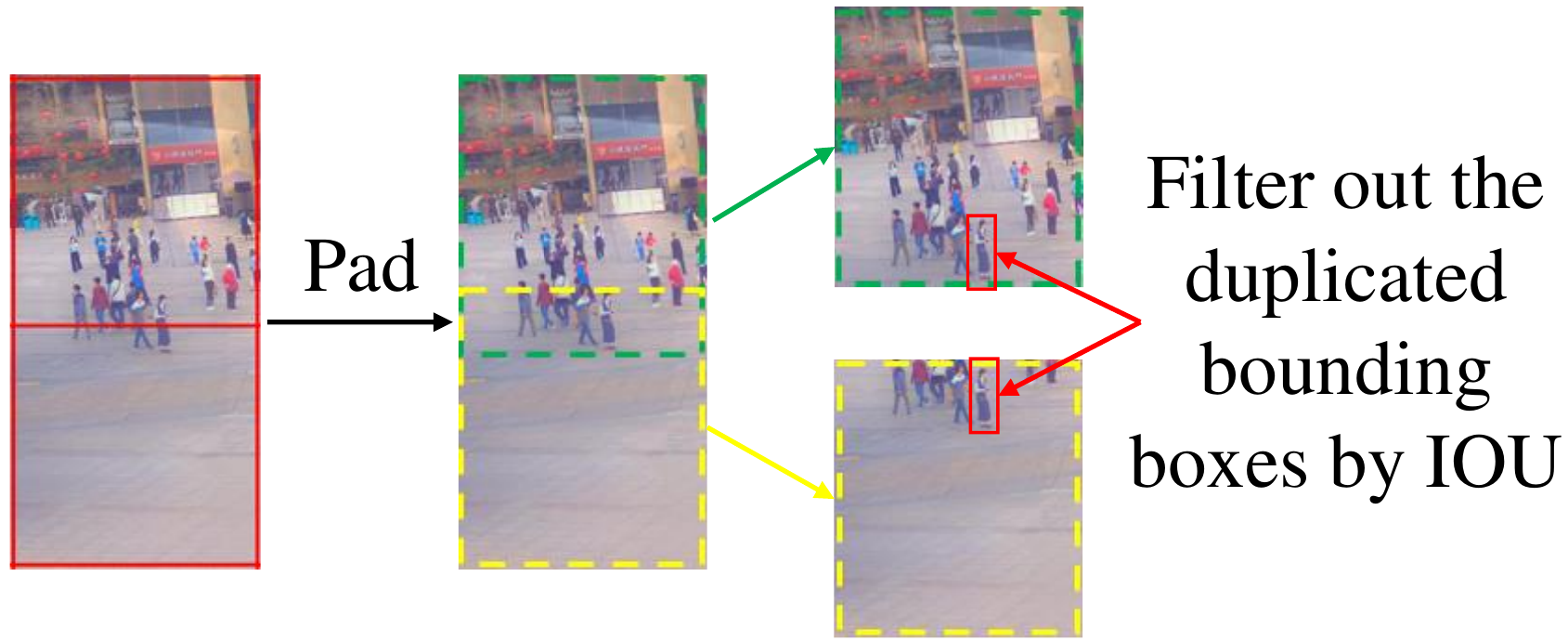}
\caption{An illustration of padding.}
\label{fig:padding}
\vspace{-0.3cm}
\end{figure}

As shown in Fig.~\ref{fig:padding}, pedestrians on the split lines will appear completely in two split regions after padding. The sizes of padding are equal to the height and width of the pedestrians. However, padding will cause the pedestrians on the split lines to be detected repeatedly. The camera can filter out the duplicated bounding boxes by intersection over union (IoU) when merging the pedestrian detection results returned by edge nodes.

After splitting and padding, the camera then puts the pedestrian detection results of the historical frames into our spatio-temporal flow filtering method to filter out the split regions that are predicted to have no pedestrians in the current frame. Next, the camera assigns the split regions to the currently available edge nodes based on our accuracy-aware DRL-based load-balanced scheduling algorithm. After receiving the split regions from the camera, edge nodes perform pedestrian detection on these split regions and return the detected bounding boxes to the camera. Finally, the camera merges the pedestrian detection results returned by edge nodes.

Then, we introduce the details of our flow spatio-temporal flow filtering method and accuracy-aware DRL-based load-balanced scheduling algorithm.

\subsection{Spatio-temporal Flow Filtering}
\begin{figure}[t]
\centering
\includegraphics[scale=0.39]{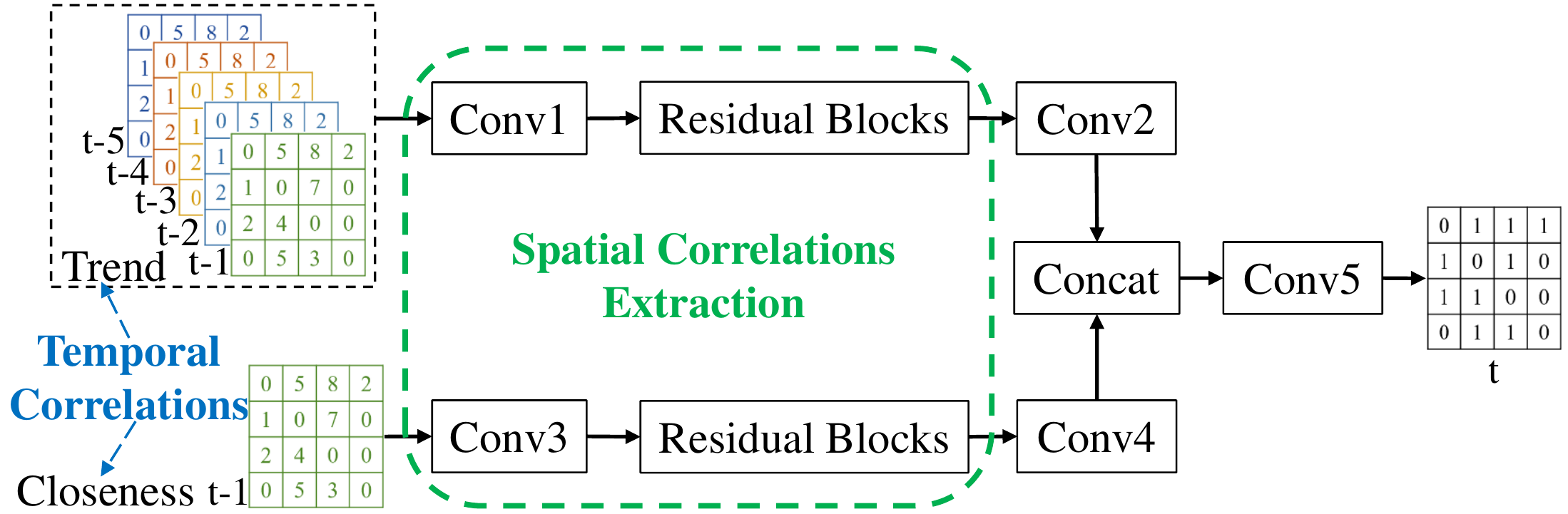}
\caption{Our spatio-temporal flow filtering model.}
\label{fig:STFF}
\vspace{-0.3cm}
\end{figure}

Fig.~\ref{fig:STFF} shows the structure of our spatio-temporal flow filtering model. Our flow filtering model is a classification model. \textsc{Hode} first transforms the pedestrian detection results of some historical frames into some matrices. The elements in these matrices represent the number of pedestrians detected in the corresponding regions. Then, \textsc{Hode} puts these matrices into the flow filtering model to filter out the small regions predicted to have no pedestrians by temporal and spatial correlations. Then, we introduce the temporal and spatial correlations used in our flow filtering model.

\textbf{Temporal Correlations:}
If a region had pedestrians in the previous several frames, it is likely that there are pedestrians in this region at the current moment. On the contrary, if a region had no pedestrians in the previous several frames, then this region is also likely to have no pedestrians at the current moment. Therefore, we can use temporal correlations to predict whether there are pedestrians in each region of the current frame. \textsc{Hode} used two kinds of temporal correlations. 

The first kind of temporal correlations is trend. The number of pedestrians may show an increasing or decreasing trend over time. So, we can put the pedestrian detection results of historical frames into a neural network to help predict whether each region in the current frame has or does not have pedestrians. As shown in Fig.~\ref{fig:STFF}, \textsc{Hode} first transforms the pedestrian detection results of the previous five frames into five matrices. Then, \textsc{Hode} puts these five matrices into a residual convolutional network to capture the trend.

The second kind of temporal correlations is closeness. \textsc{Hode} puts the pedestrian detection results at time $t-1$ into another residual convolutional network to help predict whether there are pedestrians in each region of the current frame. Because time $t-1$ is very close to the current time $t$, there is a strong temporal correlation between the number of their pedestrians. So we can use the closeness to help the prediction.

Finally, \textsc{Hode} combines the outputs of these two residual convolutional networks to filter out small regions without pedestrians.

\textbf{Spatial Correlations:}
Since spatially adjacent regions influence each other, we can use the information of adjacent regions when predicting whether a region has pedestrians or not. So, in the above two residual convolutional networks, \textsc{Hode} uses some 3×3 convolutional kernels to capture the spatial correlations between spatially adjacent regions to help filter out the small regions without pedestrians.

\textbf{Output of Spatio-temporal Flow Filtering Model:}
As shown in Fig.~\ref{fig:STFF}, the output of our spatio-temporal flow filtering model is a matrix. The elements of this matrix are either ``0'' or ``1''. ``0'' means that the corresponding region is predicted to have no pedestrians at time $t$. ``1'' means that the corresponding region is predicted to have pedestrians at time $t$. \textsc{Hode} can filter out the regions that are predicted to have no pedestrians based on this matrix. Note that here we adopt binary predictions of the pedestrian existence in regions instead of the very complicated pedestrian number prediction, in order to achieve lightweight flow filtering model deployment with high accuracy in practice.

\subsection{Accuracy-aware DRL-based Load-balanced Scheduling}
After flow filtering, the camera can assign small regions to some nearby edge nodes to perform pedestrian detection in parallel to further accelerate high-resolution pedestrian detection. However, edge nodes usually have different computing resources and the computing power is dynamically changing. In addition, different edge nodes may run different pedestrian detection models. Therefore, it is not easy to alleviate the straggler problem when assigning small regions to multiple heterogeneous edge nodes. Moreover, the accuracy of pedestrian detection also needs to be considered when assigning. Note that a high-resolution image will be partitioned into small regions of small data size for parallel edge offloading with high-speed local transmissions, and hence the networking is not considered as a bottleneck issue in this study.

We propose an accuracy-aware DRL-based load-balanced scheduling algorithm to alleviate the straggler problem and improve the accuracy of pedestrian detection. As illustrated in Fig. 7, in order to avoid complicated high-dimensional decision makings in DRL, we decompose the entire scheduling mechanism into two phases. The goal of DRL-based load-balanced scheduling phase is to balance the workload (i.e., the number of partitioned small regions assignment) among multiple heterogeneous edge nodes to alleviate the straggler problem (i.e., the completion time of one high-resolution image inference depends on the slowest edge node). The goal of accuracy-aware region dispatching phase is to improve the accuracy of pedestrian detection by a finer-grained assignment of the specific partitioned small regions to different edge nodes. We now introduce the details of these two scheduling phases.

\begin{figure}[t]
\centering
\includegraphics[scale=0.52]{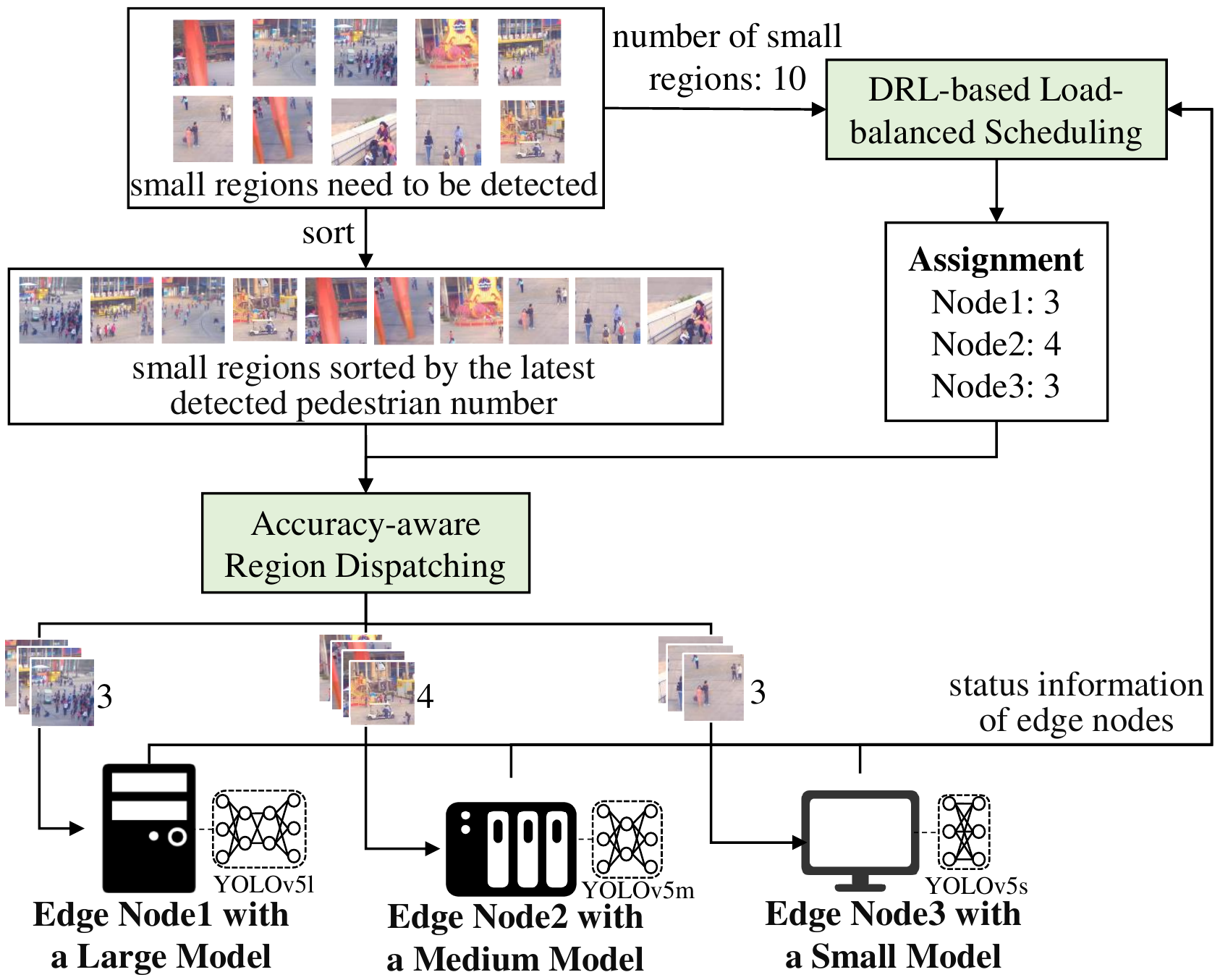}
\caption{Accuracy-aware DRL-based load-balanced scheduling.}
\label{fig:Scheduling}
\vspace{-0.3cm}
\end{figure}

\begin{itemize}
\item \textbf{DRL-based Load-balanced Scheduling:} As shown in Fig.~\ref{fig:Scheduling}, the camera first gets the current status information (specified later) of all edge nodes. Then, the camera inputs the current status information and the number of small regions into our DRL-based load-balanced scheduling algorithm. Finally, the camera can obtain the number of small regions that each edge node shall be responsible for.
\item \textbf{Accuracy-aware Region Dispatching:} After DRL-based load-balanced scheduling, our accuracy-aware region dispatching algorithm assigns specific small regions to edge nodes based on the assignment generated by DRL-based load-balanced scheduling. As shown in Fig.~\ref{fig:Scheduling}, the camera first sorts all small regions by the number of pedestrians from the latest pedestrian detection result from the previous frames, which can serve as a very fast approximate estimation of the pedestrian distributions. Then, according to the number of small regions that each edge node is responsible for, small regions with a large number of pedestrians are assigned to edge nodes using the large model, and small regions with a small number of pedestrians are assigned to edge nodes using the small model. Because a larger number of pedestrians means that the occlusion between pedestrians is more severe, it is difficult to accurately identify the pedestrians in the regions with a lot of pedestrians. Therefore, \textsc{Hode} assigns the regions with a large number of pedestrians to edge nodes using the large model to improve the accuracy of pedestrian detection.
\end{itemize}

Then, we introduce the state, action, and reward of our DRL-based load-balanced scheduling, which also implicitly takes into account the effects of accuracy-aware region dispatching among the edge nodes.

\begin{itemize}
\item \textbf{State:} We define the state at time $t$ as
\begin{equation}
s_t = (q^t_1, v^t_1, q^t_2, v^t_2, \ldots, q^t_M, v^t_M),
\end{equation}
where $M$ is the number of edge nodes, $q^t_i$ is the length of the task queue on edge node $i$ at time $t$, and $v^t_i$ is the inference speed of edge node $i$ at time $t$.

\item \textbf{Action:} The action $a_t$ taken by the camera at time $t$ is defined as
\begin{equation}
a_t = (assign^t_1, assign^t_2, \ldots, assign^t_M),
\end{equation}
where $assign^t_i$ denotes the assignment proportion of edge node $i$ at time $t$. In addition, $assign^t_i$ should meet the following constraints:
\begin{equation}
\sum_{i=1}^M assign^t_i = 1,
\end{equation}
\begin{equation}
0 \leq assign^t_i \leq 1.
\end{equation}
For simplicity, we also discretize the action space with a granularity of 0.1 and implement a Deep Q Learning (DQN) based load-balanced scheduling algorithm.

\item \textbf{Reward:} Since the goal of DRL-based load-balanced scheduling is to balance the workload among multiple heterogeneous edge nodes to alleviate the straggler problem for fast pedestrian detection, we define the reward for taking action $a_t$ in state $s_t$ as
\begin{equation}
r_t = \lambda_1 \Delta^t_p + \lambda_2 \Delta^t_q,
\end{equation}
\begin{equation}
\Delta^t_p = \sum_{i=1}^M (p^t_i - avgp^t)^2 - \sum_{i=1}^M (p^{t+1}_i - avgp^{t+1})^2,
\end{equation}
\iffalse
\begin{equation}
avgp^t = \frac{1}{M} \sum_{i=1}^M p^t_i,
\end{equation}
\fi
\begin{equation}
\Delta^t_q = \sum_{i=1}^M (\frac{q^t_i}{v^t_i} - avgq^t)^2 - \sum_{i=1}^M (\frac{q^{t+1}_i}{v^{t+1}_i} - avgq^{t+1})^2,
\end{equation}
\iffalse
\begin{equation}
avgq^t = \frac{1}{M} \sum_{i=1}^M \frac{q^t_i}{v^t_i},
\end{equation}
\fi
\iffalse
\begin{minipage}[t]{0.22\textwidth}
\begin{equation}
avgp^t = \frac{1}{M} \sum_{i=1}^M p^t_i,
\end{equation}
\end{minipage}
\begin{minipage}[t]{0.22\textwidth}
\begin{equation}
avgq^t = \frac{1}{M} \sum_{i=1}^M \frac{q^t_i}{v^t_i},
\end{equation}
\end{minipage}
\begin{equation}
\Delta^t_p = \sum_{i=1}^M (p^t_i - \frac{1}{M} \sum_{i=1}^M p^t_i)^2 - \sum_{i=1}^M (p^{t+1}_i - \frac{1}{M} \sum_{i=1}^M p^{t+1}_i)^2,
\end{equation}
\begin{equation}
\Delta^t_q = \sum_{i=1}^M (\frac{q^t_i}{v^t_i} - \frac{1}{M} \sum_{i=1}^M \frac{q^t_i}{v^t_i})^2 - \sum_{i=1}^M (\frac{q^{t+1}_i}{v^{t+1}_i} - \frac{1}{M} \sum_{i=1}^M \frac{q^{t+1}_i}{v^{t+1}_i})^2,
\end{equation}
\fi
where $\Delta^t_p$ denotes the improvement in the variance of the inference progress of edge nodes, $\Delta^t_q$ denotes the improvement in the variance of the completion time of the remaining tasks on edge nodes, $\lambda_1$ and $\lambda_2$ are two constants greater than 0, $p^t_i$ denotes the inference progress of edge node $i$ at time $t$, $avgp^t$ denotes the average inference progress of all edge nodes at time $t$, $\frac{q^t_i}{v^t_i}$ denotes the estimated completion time of the remaining tasks on edge node $i$ at time $t$, $avgq^t$ denotes the average estimated completion time of the remaining tasks on all edge nodes at time $t$. The goal of $\Delta^t_p$ is to balance the inference progress of edge nodes which is also impacted by the accuracy-aware region dispatching among the edge nodes. If the inference progress of edge nodes becomes more balanced after taking action $a_t$, then $\Delta^t_p$ will be greater than 0, indicating $a_t$ is a good action. On the contrary, if the inference progress of edge nodes becomes more unbalanced after taking action $a_t$, then $\Delta^t_p$ will be less than 0, indicating $a_t$ is a bad action. The goal of $\Delta^t_q$ is to balance the completion time of the remaining tasks on all edge nodes. Based on the rewards, DQN learns how to adjust the number of small regions assigned to each edge node to balance the progress of each node to alleviate the straggler problem.
\end{itemize}

After defining state, action, and reward, our accuracy-aware DQN-based load-balanced scheduling algorithm is presented in Algorithm \ref{alg:1}.

\begin{algorithm}[htbp]
\caption{Accuracy-aware DQN-based Load-balanced Scheduling Algorithm}
\label{alg:1}
\begin{algorithmic}[1]
\STATE Initialize the parameters of DQN and the time interval $I$ for DQN learning;
\FOR{$t$ = 0, 1, 2, \ldots}
\STATE Get the state $s_t$ at time $t$;
\STATE Put $s_t$ into DQN and choose an action $a_t$ by $\epsilon$-greedy policy;
\STATE Calculate the assignment by $a_t$ and the number of small regions;
\STATE Sort small regions by the number of pedestrians from the latest pedestrian detection result;
\STATE Send small regions to the corresponding edge nodes based on the sort and assignment;
\STATE Get the reward $r_t$ and the next state $s_{t+1}$ at time $t+1$;
\STATE Store ($s_t$, $a_t$, $r_t$, $s_{t+1}$) into the experience replay memory;
\IF {$t$ \% $I$ $==$ 0 and $t$ $>$ 0}
\STATE Randomly sample some experiences from the experience replay memory;
\STATE Learn from these experiences and update the parameters of DQN;
\ENDIF
\ENDFOR
\end{algorithmic}
\end{algorithm}

\section{Performance Evaluation}
We next introduce the setup of our experiments, the overall acceleration achieved, the performance of our flow filtering and scheduling algorithm, and the overhead of our system.

\subsection{Setup of Experiments}
We use five devices as the edge nodes, including a Dell T5820 workstation with a GTX1070 and a GTX1050 graphics card, two Jetson NXs, and one Jetson TX2. The GTX1070 graphics card is responsible for running YOLOv5m, the GTX1050 graphics card and one Jetson NX are responsible for running YOLOv5s, and the other Jetson NX and Jetson TX2 are responsible for running YOLOv5n. We use an Intel NUC 11PAH as the camera. We evaluate \textsc{Hode} on PANDA dataset \cite{PANDA} with a resolution of 4K. Cameras split 4K images by a fixed size: 512 × 512.

\subsection{Overall Performance}
We compare \textsc{Hode} with the following two methods:
\begin{itemize}
\item \textbf{Infer 4K Images:} This method assigns 4K images to edge nodes proportional to their computing power without region partitioning and flow filtering. Then, edge nodes perform pedestrian detection on these 4K images.
\item \textbf{Elf-based:} We refer to the splitting method and scheduling algorithm in Elf \cite{zhang2021elf}, and then implement an Elf-based comparison method. This comparison method first expands the bounding boxes of the latest pedestrian detection result by 30\% as the estimated location of the current pedestrians. Then this comparison method splits the estimated locations from the 4K images and assigns the split locations to edge nodes proportional to the real-time inference speed of edge nodes. Finally, edge nodes perform pedestrian detection on these locations.
\end{itemize}

\begin{figure}[t]
    \begin{minipage}[t]{0.238\textwidth}
        \includegraphics[width=1\textwidth]{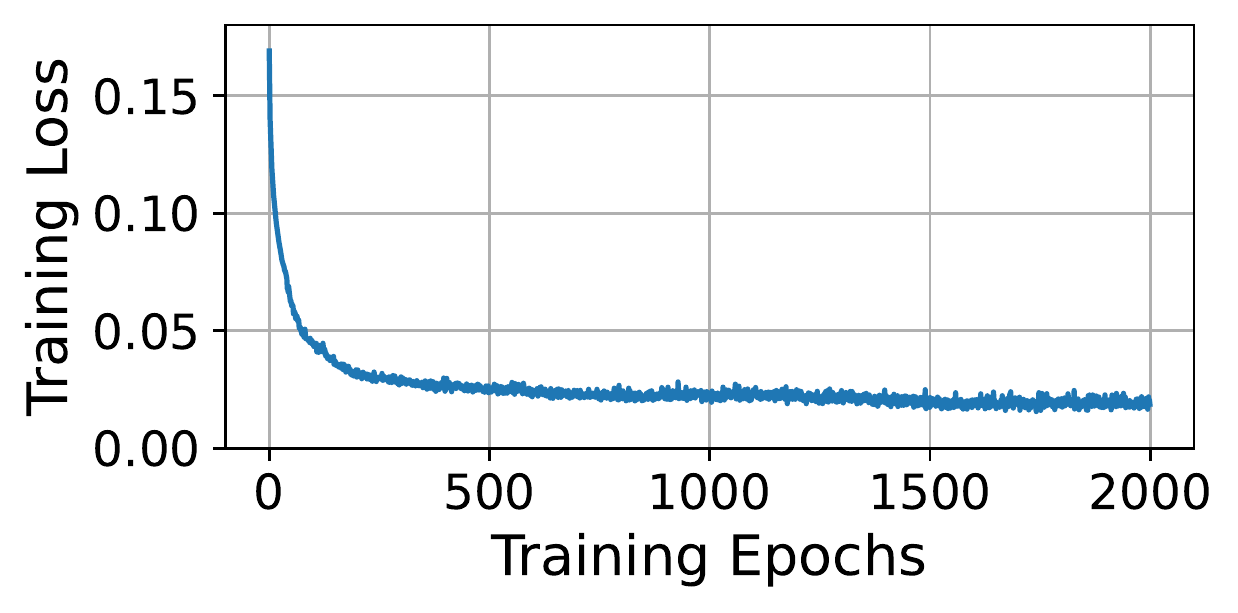}
        \caption{Training loss of our spatio-temporal flow filtering model.}
        \label{fig:STFF_loss}
    \end{minipage}
    {}
    \begin{minipage}[t]{0.228\textwidth}
        \includegraphics[width=1\textwidth]{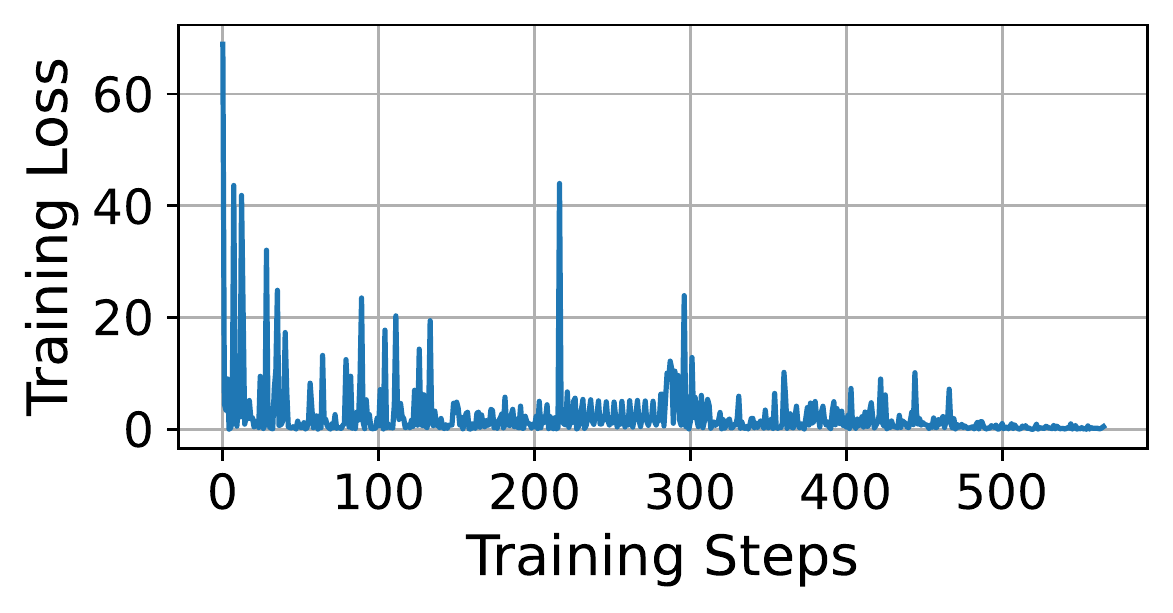}
        \caption{Training loss of our DQN-based load-balanced scheduling.}
        \label{fig:DQN_loss}
        \end{minipage}
\end{figure}

\begin{figure}[t]
    \centering
	\subfigure[Detection result of a square.]{
		\begin{minipage}[t]{0.5\linewidth}
			\centering
			\includegraphics[scale=0.04]{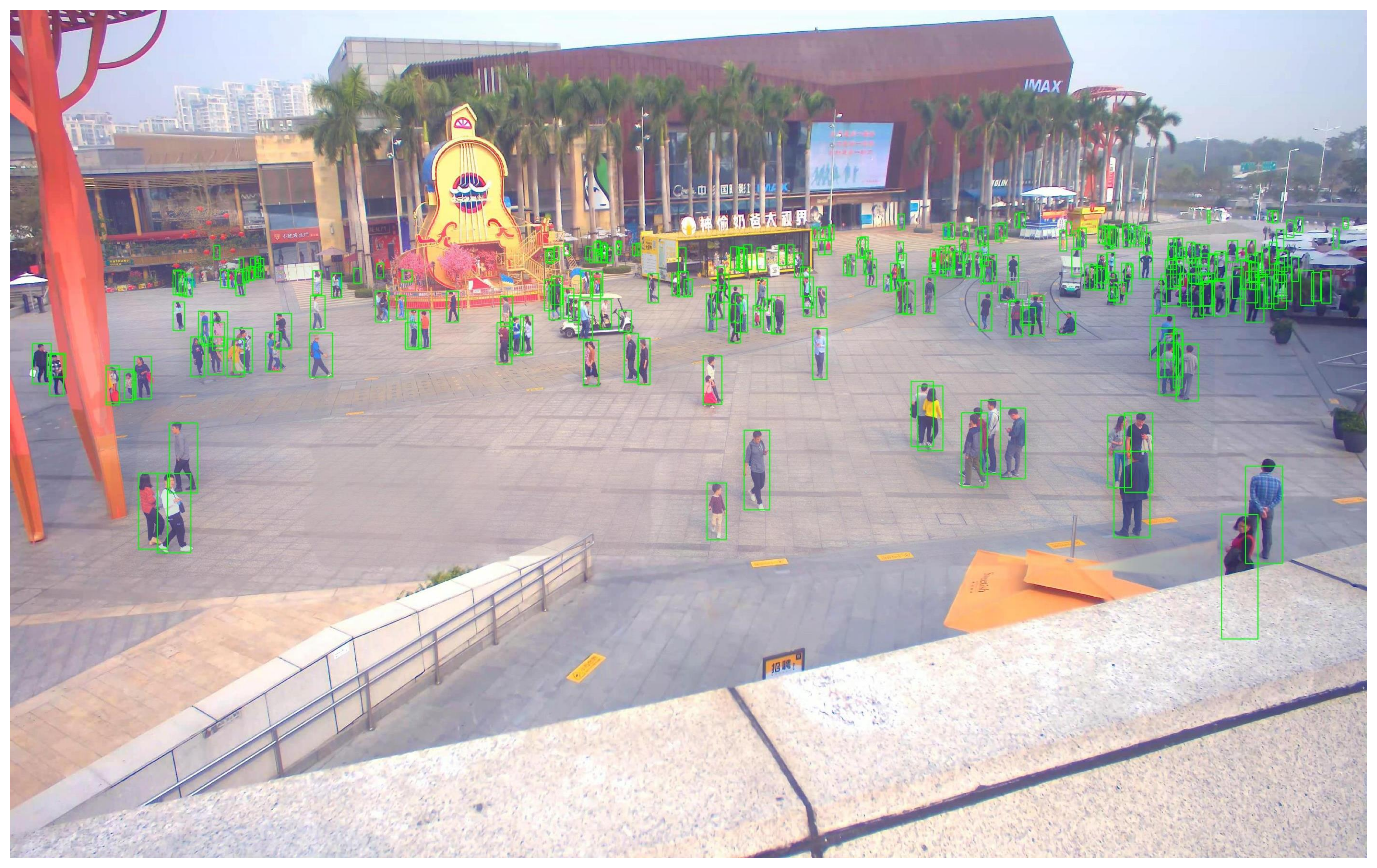}
		\end{minipage}
		\label{fig:example1}
	}%
	\centering
	\subfigure[Filtering result of (a).]{
		\begin{minipage}[t]{0.5\linewidth}
			\centering
			\includegraphics[scale=0.138]{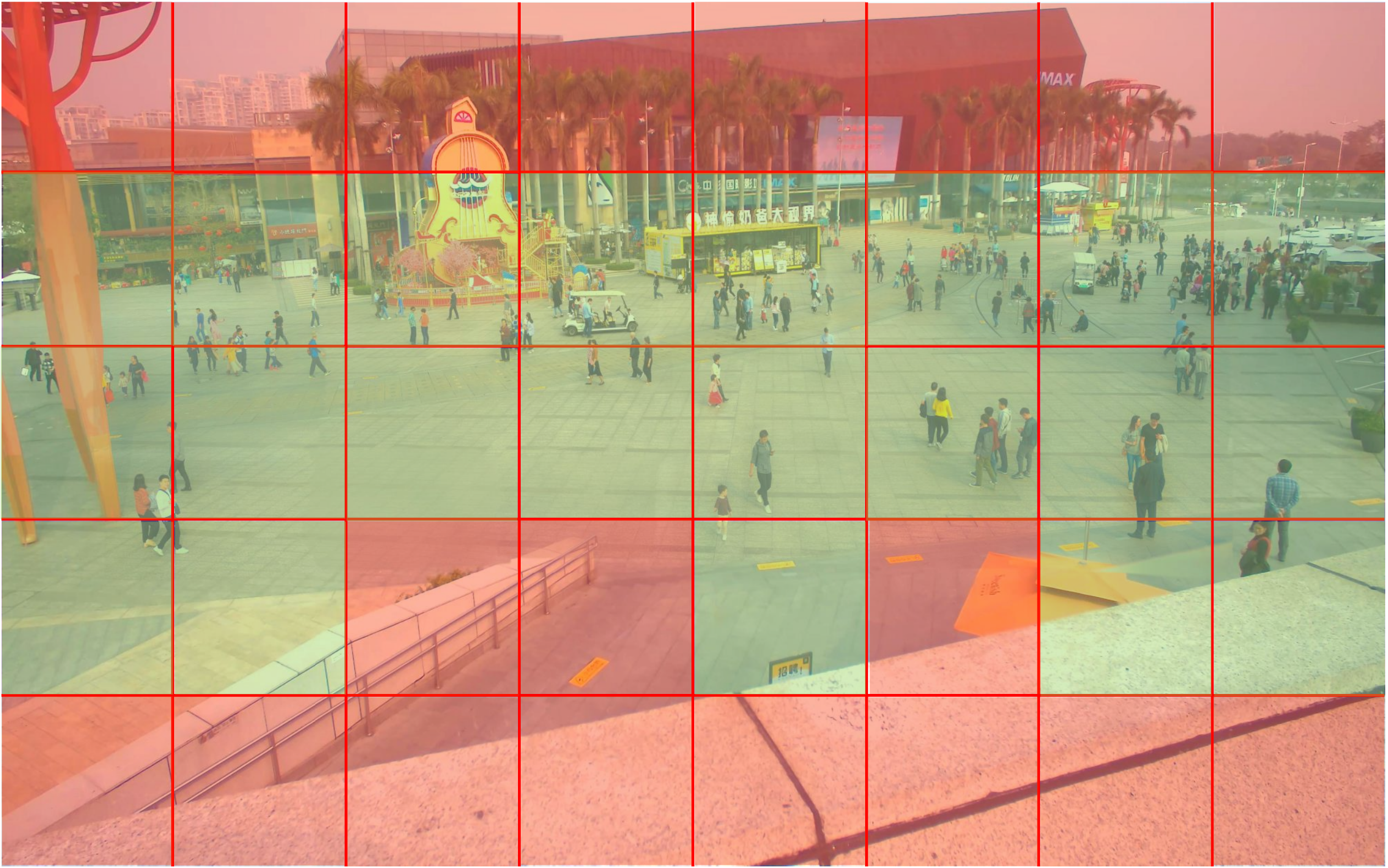}
		\end{minipage}
		\label{fig:example3}
	}\\
	\centering
	\subfigure[Detection result of a street.]{
		\begin{minipage}[t]{0.5\linewidth}
			\centering
			\includegraphics[scale=0.04]{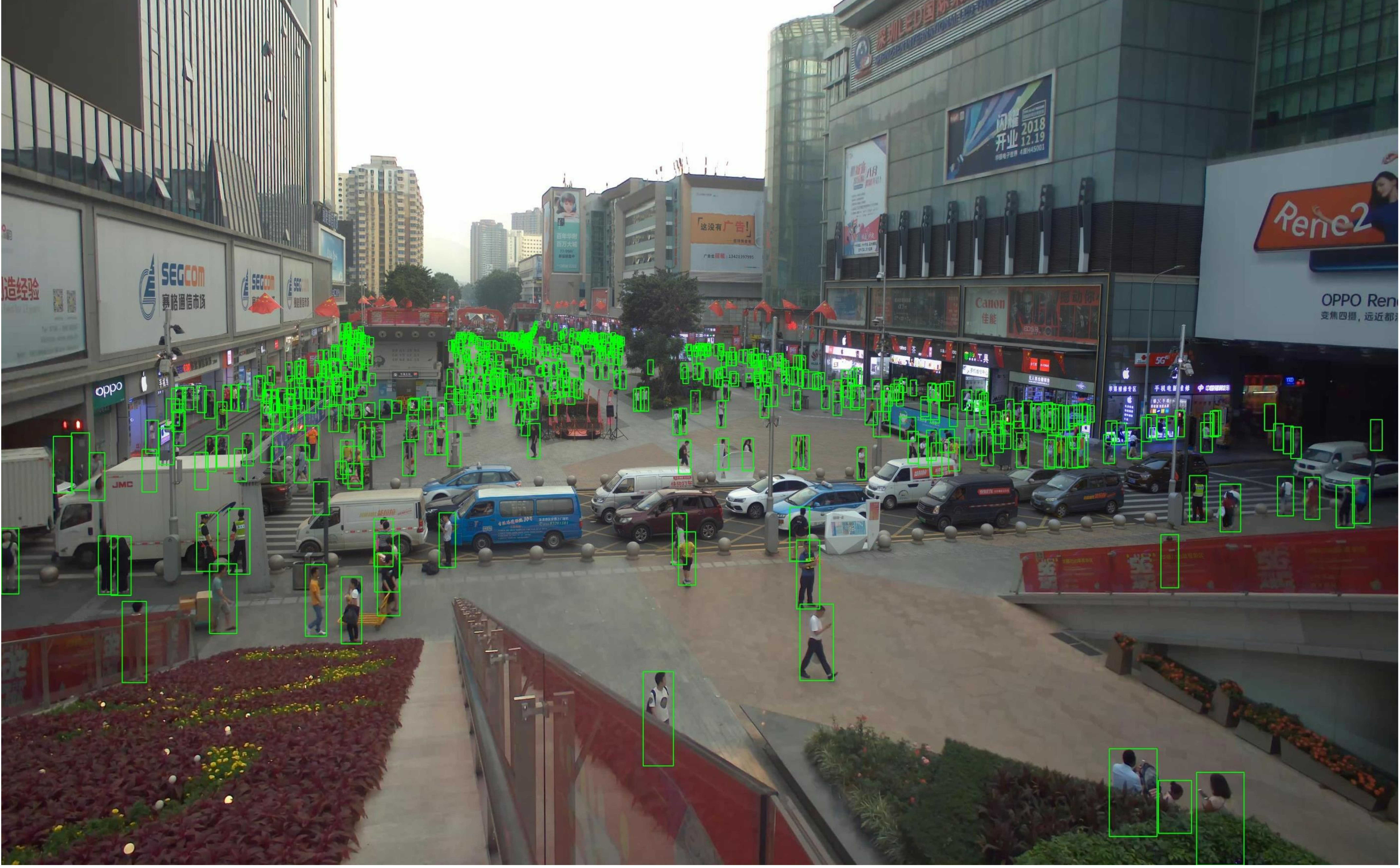}
		\end{minipage}
		\label{fig:example2}
	}%
	\centering
	\subfigure[Filtering result of (c).]{
		\begin{minipage}[t]{0.5\linewidth}
			\centering
			\includegraphics[scale=0.138]{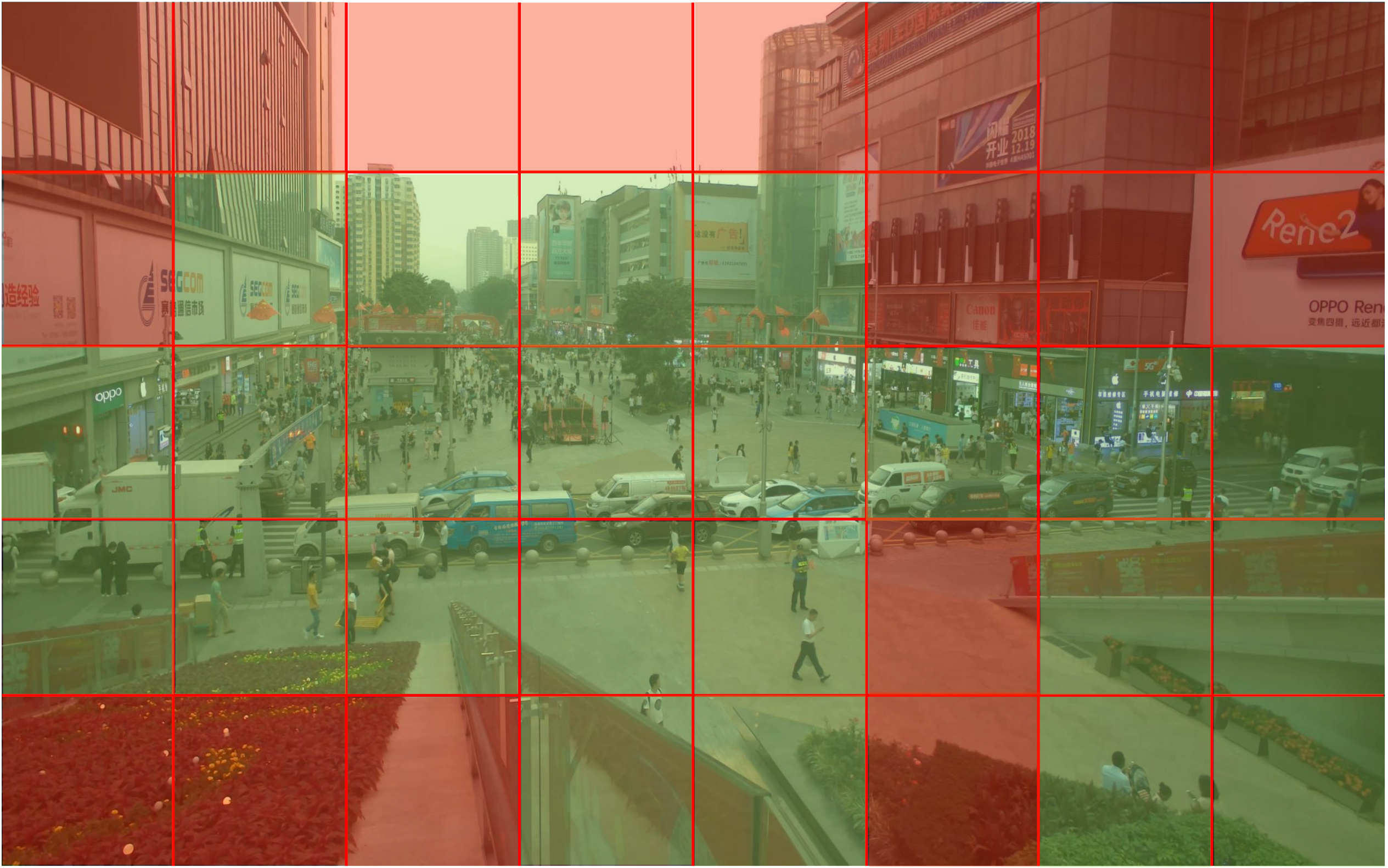}
		\end{minipage}
		\label{fig:example4}
	}%
    \caption{Some examples of \textsc{Hode}'s pedestrian detection results and filtering results in PANDA.}
    \label{fig:example_result}
\vspace{-0.5cm}
\end{figure}

Fig.~\ref{fig:example_result} shows some examples of \textsc{Hode}'s pedestrian detection results and filtering results in PANDA. Fig.~\ref{fig:Speed_mAP_evalution} shows the evaluation results of the above two comparison methods and \textsc{Hode}. Compared with inferring 4K images, \textsc{Hode} improves the inference speed from 6.02 frames per second (fps) to 12.13 fps, achieving a 2.01× speedup with less than 1\% accuracy sacrifice. In addition, the inference speed as well as mAP of \textsc{Hode} are better than Elf-based. The assignment method in Elf may lead to a lot of background pixels in the split regions, which increases the latency of inference. Moreover,  Elf did not consider assigning regions with a lot of pedestrians to the edge nodes responsible for a large model.

\subsection{Evaluation of Spatio-temporal Flow Filtering}
We divide PANDA dataset into a training set, a validation set, and a test set in a ratio of 3:1:1. Then, we use the data from the training set to train our spatio-temporal flow filtering model. We show the training loss curve of our spatio-temporal flow filtering model in Fig.~\ref{fig:STFF_loss}. Finally, we conduct some experiments to evaluate our spatio-temporal flow filtering method. The comparison method is: if a small region does not have pedestrians at time $t-i$, this small region will be filtered out at time $t$. We name this comparison method as Comp$-i$.

\begin{figure*}[htbp]
    \begin{minipage}[t]{0.32\textwidth}
        \includegraphics[width=1\textwidth]{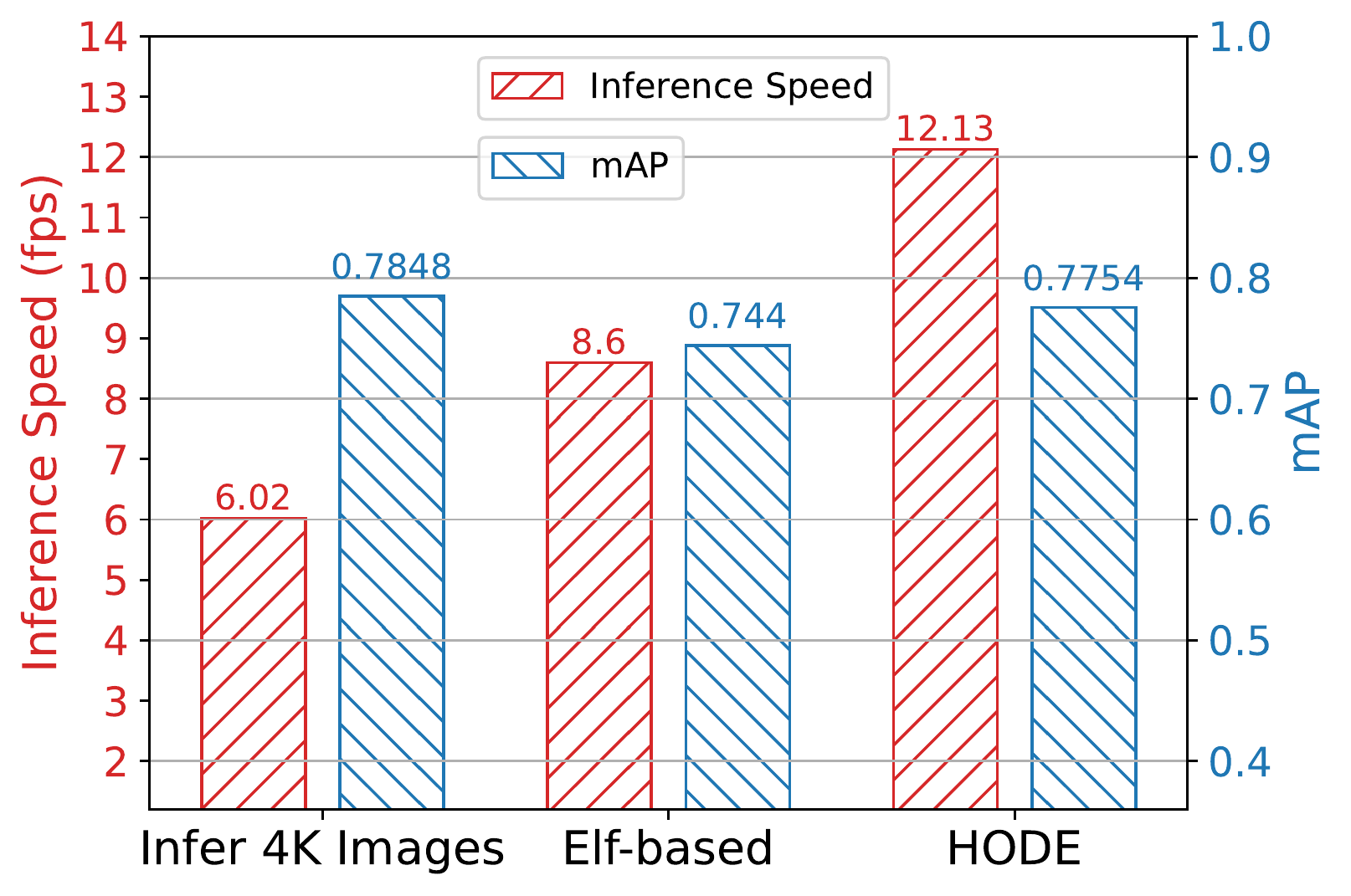}
        \caption{Inference speed and mAP of Infer 4K Images, Elf-based and \textsc{Hode}.}
        \label{fig:Speed_mAP_evalution}
    \end{minipage}
    {}
    \begin{minipage}[t]{0.32\textwidth}
        \includegraphics[width=1\textwidth]{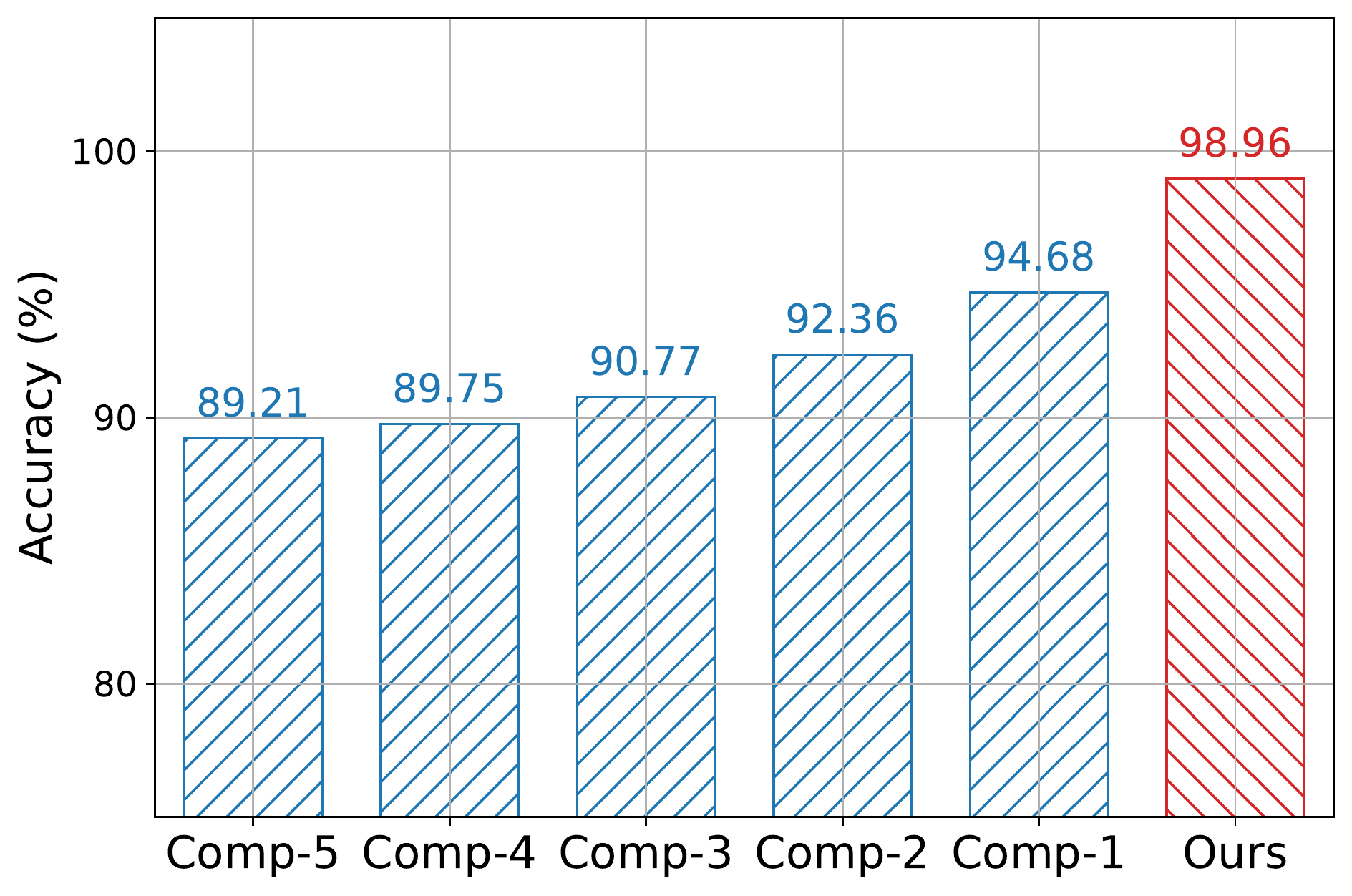}
        \caption{Accuracy of our flow filtering method and comparison methods.}
        \label{fig:STFF_evalute}
    \end{minipage}
    {}
    \begin{minipage}[t]{0.32\textwidth}
        \includegraphics[width=1\textwidth]{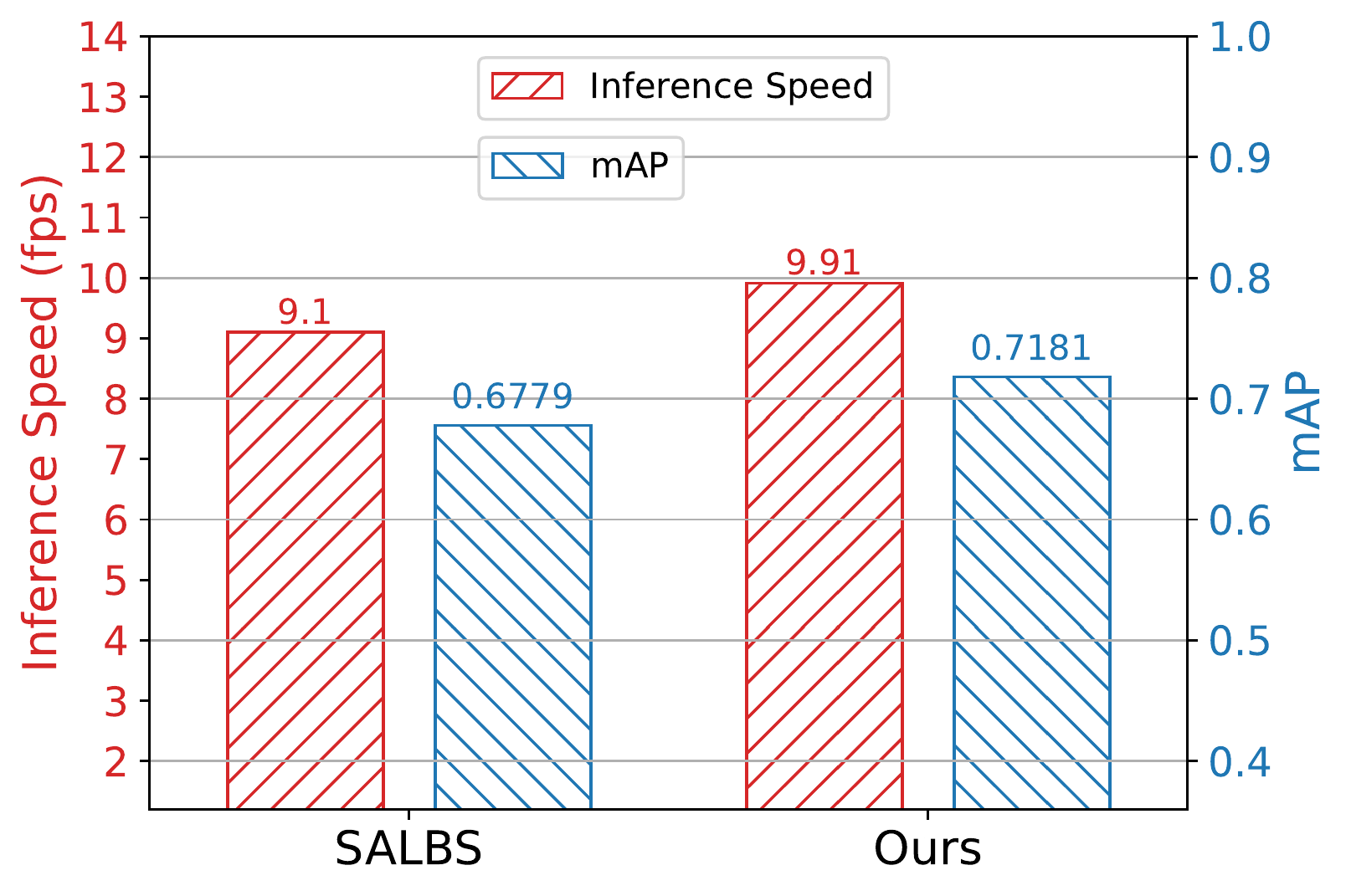}
        \caption{Inference speed and mAP of SALBS and ours.}
        \label{fig:DQN_evalute}
    \end{minipage}
\vspace{-0.2cm}
\end{figure*}

The results in Fig.~\ref{fig:STFF_evalute} show that the accuracy of our flow filtering can reach 98.96\%, which means that small regions without pedestrians can be filtered out accurately.

\subsection{Evaluation of Accuracy-aware DRL-based Load-balanced Scheduling}
Since the computing resources of edge nodes are dynamically changing in the real world, we conduct some experiments to evaluate whether our accuracy-aware DRL-based load-balanced scheduling can adapt to the dynamically changing environment.

We dynamically change the computing power of some edge nodes during the inference process. Then, we compare our accuracy-aware DQN-based load-balanced scheduling with a method that assigns small regions based on the inference speed of edge nodes in the same setting. We name this comparison method as speed-aware load-balanced scheduling (SALBS). Fig.~\ref{fig:DQN_evalute} shows the inference speed and mAP of these two methods. In contrast to SALBS, our method is more suitable for dynamic heterogeneous edge environments. We show the training loss curve of our DQN-based load-balanced scheduling algorithm in Fig.~\ref{fig:DQN_loss}.

\subsection{System Overhead}
Finally, we evaluate the system overhead produced by our flow filtering and scheduling algorithm. The latency of our flow filtering and scheduling algorithm on the camera side is 2.7ms and 1ms respectively, which is sufficiently low and suitable for resource-constrained edge nodes.

\section{Conclusion}
In order to speed up high-resolution pedestrian detection at the edge, we design \textsc{Hode}. \textsc{Hode} splits high-resolution images into some small regions and employs a spatio-temporal flow filtering method to filter out small regions without pedestrians, so as to avoid some unnecessary pedestrian detection. We also propose an accuracy-aware DRL-based load-balanced scheduling algorithm. This algorithm considers the heterogeneity of multiple edge nodes and attempts to improve the accuracy of pedestrian detection while alleviating the straggler problem. Evaluation results show that \textsc{Hode} can achieve 2.01× speedup on high-resolution pedestrian detection with less than 1\% accuracy sacrifice. In addition, our spatio-temporal flow filtering can accurately and quickly filter out regions without pedestrians before pedestrian detection.

\bibliography{ref}

\iffalse

\fi

\end{document}